\documentclass[final]{IEEEtran}

\usepackage{amsthm,amssymb,graphicx,multirow,amsmath,color,amsfonts}
\usepackage{subcaption}
\usepackage[update,prepend]{epstopdf}
\usepackage[noadjust]{cite}
\usepackage[utf8]{inputenc}
\usepackage{tikz}
\usetikzlibrary{arrows,calc}		
\usepackage{bbm} 
\usepackage{pdfpages}
\usepackage{tabulary}
\usepackage{multirow}
\usepackage{comment}
\usepackage{booktabs}

\allowdisplaybreaks 

\def\nb0{{\mathbf{0}}}
\def\nb1{{\mathbf{1}}}

\newtheorem{theorem}{Theorem}

\newtheorem{remark}{Remark}

\IEEEaftertitletext{\vspace{-15pt}}

\begin{document}
\graphicspath{{./Figures/}}
\title{
Fourier Preconditioning for Neural Feature Learning
}
\author{
Preston Pitzer, Anish Pradhan, {\em Member, IEEE,} Harpreet S. Dhillon, {\em Fellow, IEEE}
\thanks{The authors are with Wireless@VT, Department of ECE, Virginia Tech, Blacksburg, VA. Email: \{n3kos, pradhananish1, hdhillon\}@vt.edu. This material is based upon work supported by the National Institute of
Standards and Technology (NIST), U.S. Department of Commerce, through
the Public Wireless Supply Chain Innovation Fund administered by the
National Telecommunications and Information Administration (NTIA), under
award number 51-60-IF012. Any opinions, findings, conclusions, or recom-
mendations expressed in this material are those of the authors and do not
necessarily reflect the views of NIST, NTIA, or the U.S. Department of
Commerce} 
}
\maketitle
\begin{abstract}
Mutual information (MI)-inspired feature learning techniques are capable of generating low-dimensional embeddings that retain nonlinear dependence structures, but direct estimations of MI suffer from noisy probability distribution estimates in the low-data regime. 
The H-Score objective, computed from second-order statistics, provides a practical proxy metric for training feature extraction networks. 
We prove that H-Score is invariant to invertible transformations in the unrestricted functional setting, but becomes sensitive to input basis rotations under constrained approximation classes. 
Consequently, we study unitary preconditioning for H-Score networks and show that selecting an appropriate basis rotation reduces finite-width truncation error by concentrating predictive dependence into fewer dominant modes. 
We identify the fast Fourier transform (FFT) as an effective data-independent, low-cost preconditioner for approximately stationary processes, where spectral structure induces concentration of the cross-covariance singular value spectrum. 
We introduce training-free metrics based on spectral entropy and cumulative dependence energy to quantify basis suitability and predict downstream inference gains prior to network training. 
Experiments across eight multivariate datasets demonstrate that FFT preconditioning is particularly useful in resource-constrained regimes, achieving up to $50\%$ normalized mean squared error (NMSE) reduction, while the proposed metrics correlate with observed performance gains and correctly identify cases where spectral preconditioning is detrimental.
\end{abstract}

\section{Introduction} \label{sec:intro}
\IEEEPARstart{M}{achine} learning has enabled the study of complex nonlinear relationships across high-dimensional data. However, practical hardware limitations necessitate the use of low-dimensional embedding techniques that retain dependence structures while reducing computational cost. Feature extraction networks based on the MI-inspired H-Score objective use second-order statistics to generate such embeddings \cite{JMLR:v25:23-1202}. 
Theoretically, we show these objectives are invariant to invertible transformations of the input space. Practically, finite-width network effects and limited training data introduce approximation constraints that make the learned representation sensitive to the input basis. Existing work, which typically treats the input basis as fixed, focuses on network architecture rather than the structure of input data \cite{11206294}.
We propose a novel preconditioning step for H-Score networks that minimizes error associated with these finite-width effects by compacting dependence structure into a minimal set of modes. Although principal component analysis (PCA) and canonical correlation analysis (CCA) achieve optimal linear compaction under ideal covariance estimation, empirical covariance matrices are unreliable under data constraints. For approximately stationary processes, the FFT produces an approximately diagonal covariance structure, concentrating predictive dependence into a smaller number of spectral modes. Rather than assuming this property holds for the input data, we define metrics based on spectral entropy of the cross-covariance spectrum that allow for principled \emph{a priori} determination of input basis, rather than heuristic selection. 

\subsection{Prior Art}
The challenge of extracting informative features from high-dimensional data has historically been addressed through linear methods like PCA and CCA \cite{SALEM2019292}. While these methods provide optimal basis rotations for linearly dependent processes, they rely on empirical covariance estimates that become unreliable and often noise-dominated in resource-constrained regimes \cite{Gewers_2021}. Furthermore, they struggle to capture the nonlinear dependencies of interest in complex modern datasets.
To overcome these limitations, information-theoretic representation learning has emerged, leveraging surrogate objectives like the H-Score to capture nonlinear relationships through neural mappings \cite{DBLP:journals/corr/ChenDHSSA16, hjelm2019learningdeeprepresentationsmutual}. However, the transition from classical linear algebra to neural learning introduces a new bottleneck: truncation error. Because H-Score networks approximate the Hilbert-Schmidt decomposition using finite-width layers, they are most effective when the underlying dependence energy is concentrated into a small number of dominant modes.
Prior work has noted that MI estimation is significantly more efficient when dependence structures are spectrally localized or when the underlying process is approximately stationary \cite{Marques_2017, OFVERSTEDT2022196}. 
While MI in frequency has been shown to be an accurate and efficient estimator
\cite{pradhan2025mutualinformationdrivenvisualizationclustering, OFVERSTEDT2022196, Malladi_2018}, its role as a structural preconditioner to alleviate truncation error in feature learning networks remains largely unexplored.
\subsection{Contributions} \label{sec:contributions}
We show that although the H-Score objective is theoretically invariant to invertible transformations, finite-width approximation induces basis-dependent truncation error. This motivates the study of input preconditioning for practical feature learning. 
We identify the FFT as an effective fixed preconditioner for processes that are approximately stationary. To evaluate if a dataset is well suited to FFT preconditioning, we introduce \emph{training-free} spectral metrics derived from the singular value spectrum of empirical cross-covariance operators, enabling quantitative prediction of basis effects on downstream inference performance. We empirically validate the utility of this approach by measuring inference accuracy effects on FFT-preconditioned and unconditioned datasets.
\section{H-Score In Frequency} \label{sec:SysMod}

We first briefly describe how the H-Score objective follows from the goal of maximizing MI between two random variables, by maximizing the captured Hilbert-Schmidt (HS) dependence energy in the low-rank approximation of their true HS decomposition, the Canonical Dependence Kernel (CDK). We then describe how taking the Fourier transform of the input data can modify the structure of the cross-covariance, which leads the network to learn more informative dependence modes in the frequency domain. 

\subsection{A Computable Approximation of Mutual Information} \label{sec:MIF}
Consider two random variables $X$ and $Y$, which for the sake of inference may be considered as \emph{current} and \emph{future} data respectively. We compute point-wise MI (PMI) as the ratio of their joint distribution $P_{XY}$ and their marginal distributions $P_X$, $P_Y$:
\begin{align}
{\rm PMI}(x,y) :=   \log_2 \left( \frac{P_{XY}(x,y)}{P_X(x)P_Y(y)} \right).
\end{align}
These distributions are typically intractable to compute directly, so we instead look to the modal decomposition of MI as the sum of orthogonal dependence modes \cite{9174093}, which follows from the Schmidt decomposition of Hilbert spaces generated from $P_X, P_Y$ (assuming square integrability and absolute continuity of $P_XP_Y$) 
\begin{align}
P_{X,Y}(x,y)=P_X(x)P_Y(y)\left(1+\sum_{i=1}^{K}\sigma_if_i^*(x)g_i^*(y)\right).
\end{align}
From this decomposition we characterize the statistical dependence between $X$ and $Y$ using the CDK $i_{X;Y}$ \cite{JMLR:v25:23-1202}:
\begin{align}
i_{X;Y} = \sum_{i=1}^{K}\sigma_i (f_i^* \otimes g_i^*),
\end{align}
where $\|i_{X;Y}\|_{HS}=0$ implies independence of $X$ and $Y$. Just as the true distributions on $X$ and $Y$ are incomputable, the true full-rank decomposition of $i_{X;Y}$ is equivalently unknowable. We then seek a metric for computing the approximation error of a lower-rank subset of orthogonal modes. This metric, proposed in \cite{8979352} is known as H-Score ($\mathcal{H}$)
$$
\mathcal{H}(f,g):=\mathbb{E}[f^{\intercal}(X)g(Y)] - (\mathbb{E}[f(X)])^\intercal \mathbb{E}[g(Y)]-\frac{1}{2}\cdot \text{tr}(\Lambda_f\Lambda_g)
$$
$$
\Lambda_f=\mathbb{E}[f(X)f^\intercal(X)] \quad \Lambda_g = \mathbb{E}[g(Y)g^\intercal(Y)].
$$
This metric has the advantage of being computable with training data, since the expected value can be computed as the empirical mean \cite{JMLR:v25:23-1202}.

\subsection{Objective Equivalence} \label{sec:hinvariance}
We observe the effect (or lack thereof) of preconditioning input data on the H-Score objective in the full-rank case.
\begin{theorem} \label{thm:HInvaraince}
Let $T:\mathbb{R}^{d_x} \rightarrow \mathbb{R}^{d_x}$ be invertible. Suppose the function class $\mathcal{F}$ is closed under composition with $T$ and $T^{-1}$. Then
\begin{align}
    \sup_{f,g\in\mathcal{F}}\mathcal{H}(f(X),g(Y))= \sup_{f,g\in\mathcal{F}}\mathcal{H}(f(TX),g(Y)).
\end{align}
\end{theorem}
The proof follows from bijectivity of composition. Because $T$ is invertible and $\mathcal{F}$ is closed under composition with $T$ and $T^{-1}$, the map $f\rightarrow f \circ T$ is a bijection on $\mathcal{F}$. Hence the supremum over $\mathcal{F}$ is invariant under a change of variables $X \rightarrow TX$
\begin{remark}
    This invariance does not hold practically for restricted function classes, like those that appear in finite-width neural networks, which explains empirically observed performance differences. 
\end{remark}
\subsection{Basis Transformation and Low-Rank Approximation}
\label{sec:dft}
We now formalize the restriction finite-width neural networks impose on the HS Decomposition. First, let $\mathcal{H}_X = L^2(P_X)$ and $\mathcal{H}_Y = L^2(P_Y)$ be the Hilbert spaces of square-integrable functions of $X$ and $Y$. The CDK $i_{X;Y}$ is a member of tensor product space $\mathcal{H}_X \otimes \mathcal{H}_Y$. Then the approximation manifold $\mathcal{F}_k$ is:
\begin{align}
\mathcal{F}_k := \left\{ \sum_{i=1}^k w_i (\psi_i \otimes \phi_i) \mid \psi_i \in \mathcal{V}_X, \phi_i \in \mathcal{V}_Y, w_i \in \mathbb{R} \right\},
\label{eq:manifold}
\end{align}
where $\mathcal{V_X} \subset \mathcal{H}_X$ and $\mathcal{V_Y} \subset \mathcal{H}_Y$ are the functional subspaces representable by the finite-width network.
To understand the effect of basis rotations on decompositions that lie within the manifold, we first consider their effect on the original tensor-product space. Let $X \in \mathbb{R}^{L \times d_x}$ and $Y \in \mathbb{R}^{L \times d_y}$ be centered random variables with cross-covariance $C_{XY}=\mathbb{E}[X^\intercal Y]$. Consider a generic unitary transformation $U \in \mathbb{C}^{L \times L}$ acting on the input space such that $X_U = UX$. The transformed cross-covariance follows as 
\begin{align}
C_{X_U Y} = \mathbb{E}[(UX)^\intercal Y] = UC_{XY}.
\end{align}
Because $U$ is unitary, the HS norm remains invariant, which implies that the total dependence energy is preserved. However, the alignment of the operator's singular values relative to $\mathcal{F}_k$ is not preserved.
\begin{remark}(Basis-Dependent Approximation):
Let $\mathcal{F}_k$ be a class of rank-k operators. While the singular values $\sigma_i$ of $C_{XY}$ are invariant under $U$, the projection error $\|C_U -  P_{\mathcal{F}_k} \|_{HS}$ depends on how well $U$ aligns the dominant singular values with directions favored by the approximation class $\mathcal{F}_k$.
\end{remark}
In H-Score networks, projection into the approximation manifold $P_{\mathcal{F}_k}$ is truncation to the $k$ highest energy modes. Therefore an ideal $U$ maximizes energy concentrated in these $k$ modes, minimizing truncation error.
The DFT, defined as a unitary matrix $F \in \mathbb{C}^{L \times L}$, is a specific case of this transformation. We define the frequency domain cross-covariance $C_{X_f Y} = FC_{XY}$. When the underlying process is approximately stationary, the DFT is known to approximately diagonalize the covariance \cite{Marques_2017}, which aligns the dominant singular values of $C_{X_f Y}$ in a way that minimizes projection error at minimal computational cost.

\section{Basis Selection}
\label{sec:Metrics}

Although domain knowledge may suggest the use of a particular basis for low-rank projection, a numerical approach may identify more subtle dependence structures. We evaluate the effect of basis rotations on properties of the singular value spectra by using representative samples of historical data. We cover both a generic and FFT-specific method for quantifying these effects to facilitate a comparative basis selection process.

\subsection{Whitened Cross-Covariance Operator}

Let $X \in \mathbb{R}^{N \times d_x}$ and $Y \in \mathbb{R}^{N \times d_y}$ denote centered observations. The cross-covariance matrix is defined as $C_{XY} = \frac{1}{N} X^\top Y$. To ensure scale invariance, we analyze the whitened cross-covariance operator:
\begin{align}
    \widetilde{C}_{XY} = C_{XX}^{-1/2} \, C_{XY} \, C_{YY}^{-1/2},
\end{align}
where $C_{XX} = \frac{1}{N}X^\top X$ and $C_{YY} = \frac{1}{N}Y^\top Y$. The singular values of $\widetilde{C}_{XY}$, denoted $\sigma_{t,i}$, represent the energy associated with dependence structures in the temporal domain. Practically, these quantities are computed using low-rank PCA approximations for numerical efficiency.

\subsection{Frequency-Resolved Cross-Covariance}

For temporal data $X \in \mathbb{R}^{N \times L \times d_x}$, the unitary DFT applied along the time axis yields $X_f = FX$. The frequency-resolved cross-covariance at frequency $\omega$ is defined as:
\begin{align}
    C_{\omega} = \frac{1}{N} X_f(\omega)^{\dagger} Y.
\end{align}
The predictive energy at each frequency is the squared Frobenius norm $\sigma_{\omega} = \|C_{\omega}\|_F^2$. This energy distribution across frequencies provides a computable proxy for the singular value decay of the full operator $C_{X_f Y}$ described in Section \ref{sec:dft}.
\vspace{-5pt}
\subsection{Spectral Metrics for Basis Evaluation}

To provide a principled criterion for basis selection, we derive comparative statistics from the singular value spectra computed in the temporal and Fourier domains.

We quantify the concentration of predictive information by treating the normalized singular values (or frequency energies) as a probability distribution, $p_i = \sigma_i / \sum \sigma_j$. The normalized spectral entropy is:
\begin{align}
    H = -\frac{1}{\log n} \sum_{i=1}^{n} p_{i} \log p_{i},
\end{align}
where $n$ denotes the number of components. We denote the entropy in the temporal and frequency domains as $H_t$ and $H_f$.

To quantify the impact of spectral preconditioning, we define the Entropy Ratio:
\begin{align}
    \mathrm{ER} = \frac{H_f}{H_t}.
\end{align}
Values of ER $<$ 1 indicate $C_{X_f Y}$ has a more concentrated singular value spectrum than
$C_{XY}$. From Remark 2, we know that this corresponds with a reduction in finite-width
approximation error, and practically, more accurate prediction.

For H-Score network parameterization, we define the Cumulative Energy Rank as the minimum integer $k$ required to retain a threshold $\alpha$ of the total energy:
\begin{align}
    \text{CER}_\alpha = \min \left\{ k \in \mathbb{Z} : \frac{\sum_{i=1}^k \sigma^2_i}{\sum_{i=1}^K \sigma^2_i} \geq \alpha \right\}.
\end{align}
This serves as a heuristic estimate for the latent dimension of the feature extractor. Additionally, the Effective Rank, computed via the participation ratio \cite{PhysRevA.111.052614} or as the exponential of the entropy \cite{Bunea_2015}, provides a real-valued approximation for the smallest viable latent dimension.

\section{Numerical Results} \label{sec:NumResults}
\subsection{Case Study Datasets}
We evaluate the proposed framework on eight public multivariate datasets spanning traffic, solar power, weather, wireless networking, cloud gaming, and tabular regression tasks \cite{lai2018modelinglongshorttermtemporal,11206294,oranvideostreaming,graff:hal-03421031, jenaclimate}. The datasets exhibit varying levels of temporal and spectral structure, allowing evaluation of when FFT preconditioning improves low-rank feature learning. The wine quality datasets \cite{wine_quality_186} serve as a negative control due to the lack of temporal dependence.

\begin{figure*}[t]
\begin{subfigure}{\textwidth}
    \begin{center}
    \includegraphics[width=.85\linewidth]{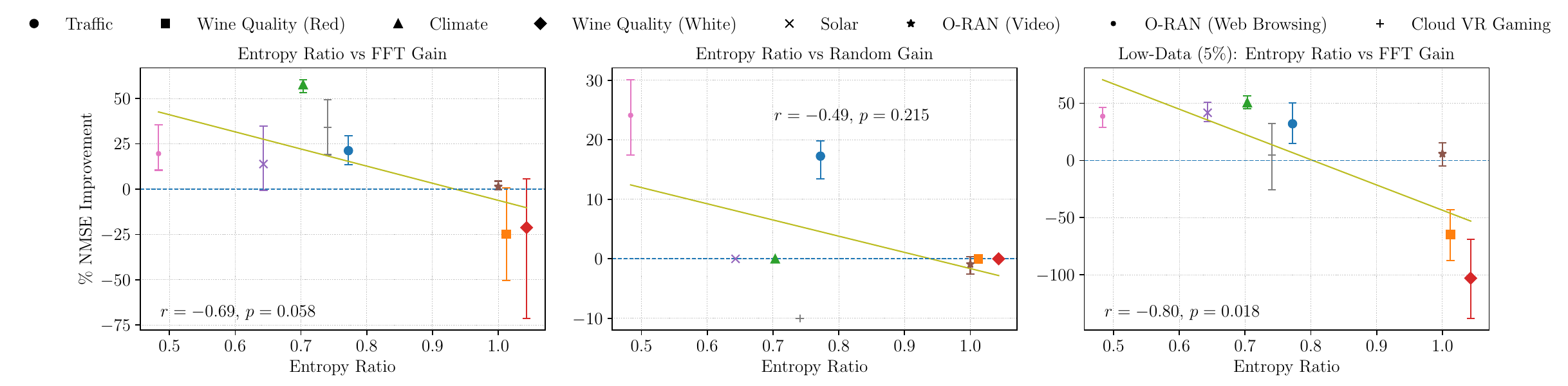}
    \end{center}
\end{subfigure}
\caption{Numeric evaluation of entropy ratio prediction accuracy}
\label{fig:numericeval}
\end{figure*}

\subsection{Experimental Design}
We evaluate the relationship of Fourier Score with Mean Squared Error (MSE) reduction associated with preconditioning the input using the FFT.
The input time series is segmented into sliding windows of length $L=29$. For each window, the first $L-1=28$ samples constitute the input $X\in \mathbb{R}^{28\times d_x}$ and the final sample is the prediction target $Y\in \mathbb{R}^{d_y}$. 
The choice of $L$ is a trade-off between FFT-induced spectral leakage effects, and retaining temporal information. Leakage leads to a higher entropy singular value spectrum by spreading energy over many bins. This effect can be mitigated by making $L$ arbitrarily large, but this in turn destroys temporal information and increases processing time.
The fence effect, which also arises from limited FFT resolution, can split energy from a dominant mode across adjacent bins when the true modal frequency is not represented by the discrete FFT grid; unlike leakage, this redistribution is localized and therefore has a smaller effect on retained energy.
For each dataset, we compute the dataset-level metrics, including the spectral entropy and $\text{CER}_{95}$ in the temporal and frequency domains. These are evaluated prior to model training and are used for correlation analysis. 
The feature extraction model consists of mappings $f_\theta(\cdot)$ and $g_\phi(\cdot)$, trained jointly to maximize H-Score. The $f$ network is an Echo State Network (ESN) with ReLU activations, and the $g$ network is an MLP with two hidden layers (16 and 32 neurons) with a ReLU activation function. \newline
The ESN may initially appear to be an odd choice when considering frequency domain inputs, as the reservoir is intended to learn temporal dependencies by considering previous input samples. However, we can exploit this structure in the frequency domain by using the reservoir to capture inter-frequency coupling, which substantially expands the learnable dynamics to include harmonic relationships, coupling factors, and the underlying power spectral density function of the data by presenting each frequency bin in monotonic order. \newline
The dimension of the output layer for each network is fixed at 8, which functionally caps the number of learnable dependence modes. The feature mapping network trains with a learning rate of 0.005 for 50 epochs, with a batch size of 128. \newline
Following training, the learned feature representation $f_\theta(X)$ is used as the input to a downstream low-dimensional MLP. This predictor is trained on MSE loss to estimate the target $Y$. 
Model performance is evaluated on data retained from train-test splits of $[0.05, 0.3, 0.6, 0.95]$. Performance is quantified as the NMSE of the inference, evaluated using either no basis rotation, a random basis rotation, or the FFT with real and imaginary parts concatenated. We present the mean and 95\% confidence interval across 10 random seeds. 
\begin{table}
\centering
\renewcommand{\arraystretch}{1.15}
\begin{tabular}{l l l l l}
\toprule
    Dataset & $\text{CER}^{(\text{freq})}_{0.95}$ & $\text{CER}^{(\text{time})}_{0.95}$ & $H_f$ & $H_t$\\ \midrule
    O-RAN 0 (Video) & 14 & 11 & 0.98 & 0.98 \\
    O-RAN 1 (Web Browsing) & 4 & 8 & 0.44 & 0.91 \\ \hline
    Cloud VR Gaming & 2 & 14 & 0.60 & 0.81 \\
    Traffic & 7 & 26 & 0.72 & 0.92 \\ 
    Wine Quality (Red) & 9 & 5 & 0.97 & 0.96 \\
    Wine Quality (White) & 8 & 9 & 0.97 & 0.93 \\ 
    Weather & 2 & 8 & 0.63 & 0.90 \\
    Solar & 2 & 19 & 0.55 & 0.86 \\ 
    \bottomrule
\end{tabular}
\caption{Structural metrics for case study datasets}
\vspace{-15pt} 

\end{table}
\vspace{-15pt}
\subsection{Spectral Transformation Results}
\begin{figure}[t]
    \centering
   \vspace{-13pt} 
   \includegraphics[width=0.7\linewidth]{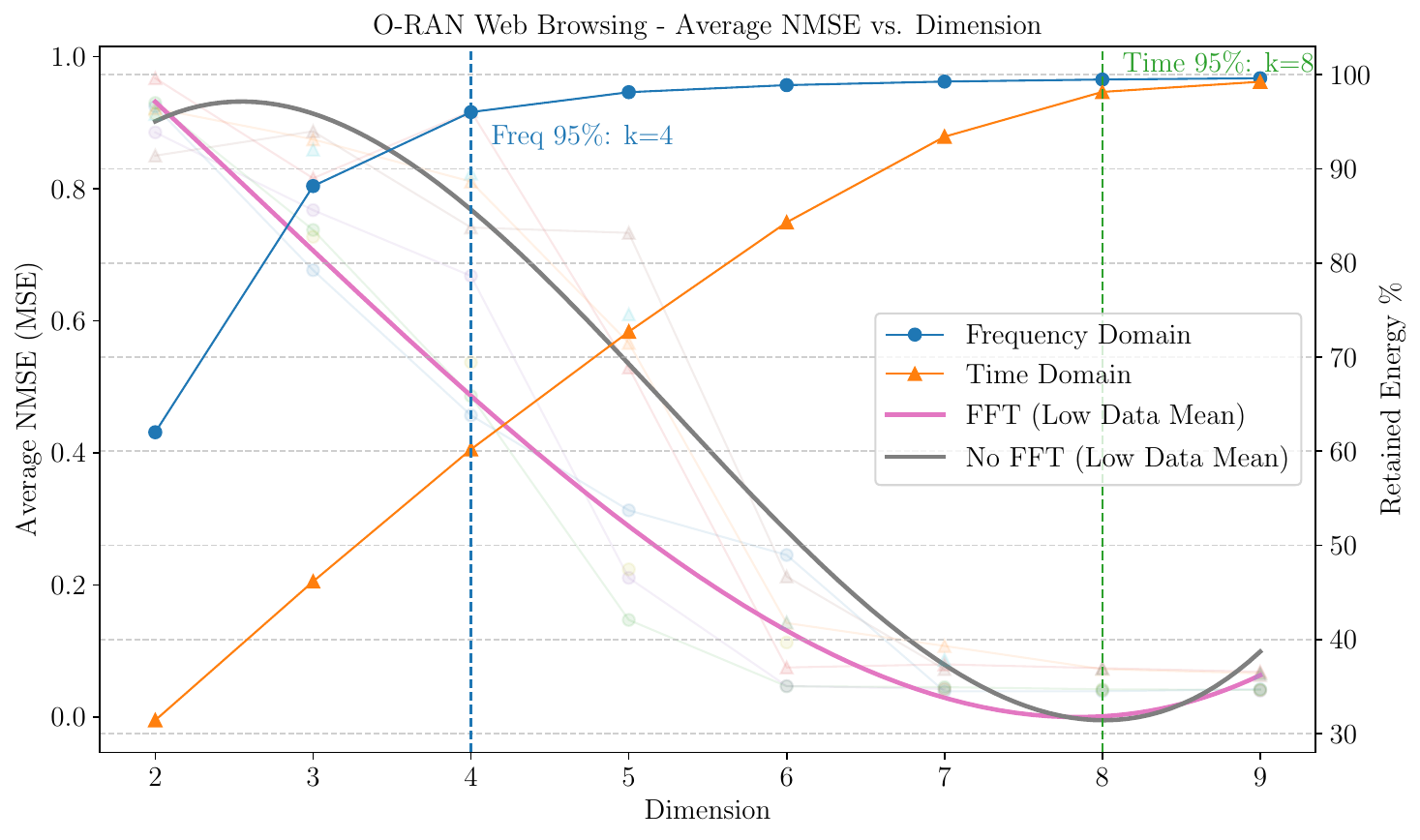}
    \vspace{-8pt}  
    \caption{Multi-domain average NMSE trend (full opacity) and per-seed (reduced opacity) NMSE vs. latent dimension}
    \label{fig:latentdim}
    \vspace{-15pt} 
\end{figure}

\vspace{-1pt}
The results in Fig.~\ref{fig:numericeval} show a statistically significant relationship between ER and NMSE improvements in the time and frequency domains in the low-data (5\% training split) case, and a near-significant relationship in the unrestricted data case. The difference between frequency and time-domain conditions is especially stark in the low-data regime, with a noticeable decline in metric accuracy and performance improvement when data availability is unrestricted. The randomized rotation condition shows a near-zero insignificant ($p=0.12$) improvement over no rotation, reinforcing the need for numeric evaluation of dependence structures before arbitrarily applying a rotation. 
The aperiodic ``wine'' datasets demonstrate a case where the metric accurately predicts a negative consequence of spectral pre-conditioning, further emphasizing the need for metric computation prior to using the FFT. 
In absolute terms, we observe NMSE decreases by as much as 50\% when minimal data is available for training. Smaller margins in the higher-data regime are expected, as any differences between function-space mappings are learned by the supervised model given sufficient data, eventually resulting in similar outcomes.
Fig.~\ref{fig:latentdim} shows the relationship between empirically observed NMSE differences across dimensions and domains for the Web Browsing dataset. The results of three random seeds are shown in low opacity, with a smoothed trend to indicate the average result. 
Retained energy, acting as a proxy for truncation error, appears predictive of inference accuracy across latent dimensions. CER, with an appropriate choice of $\alpha$, may be useful for dataset-level cost-accuracy tuning. 

\vspace{-9pt}
\section{Conclusion}
This letter identified input preconditioning as a method for optimizing H-Score feature extraction networks. We showed how the basis dependence of these networks is a result of finite-width network effects, and developed proxy metrics for evaluating the impact of a chosen basis on the resulting truncation error. The FFT appears to be a strong choice for such a preconditioner in theory, as it produces a concentrated basis without data dependence and for a minimal computational cost. Small window lengths, necessary for rapid FFT computation, also introduce spectral leakage. This leakage increases spectral entropy, degrading performance, but can be mitigated using windowing methods like Hann windowing. 
Numeric results validate both the magnitude of the preconditioning effect and the accuracy of the proxy metrics, measured using NMSE of a downstream inference model. Results in the low-data regime show a stronger effect, enhancing the already strong low-data performance of H-Score networks \cite{11206294}.
Further, we found that CER is weakly predictive of the optimal latent dimension using O-RAN data as a case study. Although the frequency domain input representation produced a more accurate inference model in highly compressed representations, the empirically optimal dimension remained the same in both domains. 
Future work may consider studying optimal latent dimension prediction, preconditioning alternative network architectures or window length effects.

\bibliographystyle{IEEEtran}
\bibliography{IEEE_Paper_Template/hokie}

\end{document}